\pdfoutput=1

\documentclass[prb,aps,twocolumn,amsmath,amssymb,floatfix,superscriptaddress]{revtex4}

\usepackage{color}
\usepackage{soul}
\usepackage{amsmath}
\usepackage{latexsym}
\usepackage{amssymb}
\usepackage{mathrsfs}
\usepackage{graphics,epstopdf}
\usepackage[colorlinks=true, citecolor=blue, urlcolor=blue ]{hyperref}
\usepackage{epsf,graphics,graphicx}

\date{\today}

\begin{document}
\title{Localization transitions in an open quasiperiodic ladder}

\author{Suparna Sarkar}
\email{suparna@jncasr.ac.in}
\thanks{These authors contributed equally to this work.}
\author{Soumya Satpathi}
\thanks{These authors contributed equally to this work.}
\author{Swapan K. Pati}
\email{pati@jncasr.ac.in}
\affiliation
{Theoretical Sciences Unit, School of Advanced Materials (SAMat), Jawaharlal Nehru Centre for Advanced Scientific Research, Bangalore 560064, India}
\date{\today} 

\begin{abstract}

We investigate localization transition in an open quasiperiodic ladder where the quasiperiodicity is described by the 
Aubry-Andr\'{e}-Harper model. While previous studies have shown that higher-order hopping or constrained quasiperiodic 
potentials can induce a mixed-phase zone in one dimension, we demonstrate that the dissipation can induce mixed phase zone in a 
one dimensional nearest-neighbor system without imposing any explicit constraints on the quasiperiodic potential or 
hopping parameter. Our approach exploits an exact correspondence between the eigenspectrum of the Liouvillian superoperator 
and that of the non-Hermitian Hamiltonian, valid for quadratic fermionic systems under linear dissipation. Using third 
quantization approach within Majorana fermionic representation, we analyze two dissipation configurations: alternating 
gain and loss at every site, and at alternate sites under balanced and imbalanced conditions. By computing the inverse 
and normalized participation ratios, we show that dissipation can drive the system into three distinct phases: delocalizd, 
mixed, and localized. Notably, the mixed-phase zone is absent for balanced dissipation at every site but emerges upon 
introducing imbalance, while for alternate site dissipation it appears in both balanced and imbalanced cases. Furthermore, 
the critical points and the width of the mixed-phase window can be selectively tuned by varying the dissipation strength. 
These findings reveal that the dissipation plays a decisive role in reshaping localization transitions in quasiperiodic systems, 
offering new insight into the interplay between non-Hermitian effects and quasiperiodic order.
\end{abstract}

\maketitle

\section{Introduction}

The phenomenon of localization in quantum systems has long been a subject of fundamental and practical interest since its 
prediction by P. W. Anderson in 1958~\cite{anderson}. The pioneering work of Anderson demonstrates that  in low-dimensional 
systems with uncorrelated site potentials, all energy eigenstates exhibit exponential localization, regardless of the disorder 
strength. It makes the system quite trivial one as no mobility edge (ME) exist to separate a conducting region from an insulating 
one, which is essential for a disorder-induced metal-to-insulator  transition~\cite{loc1,loc2}. However, In three-dimensional 
systems, disorder enables the coexistence of extended and localized states, separated by a ME at a critical energy~\cite{loc3}. 
In contrast to randomly disordered systems, a quasiperiodic system can exhibit a ME through a controllable metal-insulator 
phase transition even in low dimensions\cite{cor1,cor2}. A famous example is the Aubry-Andr\'{e}-Harper (AAH)~\cite{aah1,aah2} 
model, in which the on-site potentials follow a cosine modulation pattern. It exhibits a sharp transition where all eigenstates are 
delocalized below a critical point but become fully localized beyond it. Since the critical disorder strength does not depend on 
eigenenergy, there is no ME in such systems~\cite{aah2}. The emergence of mobility edge requires coupling of two 
one-dimensional AAH chains to construct a ladder, along with the presence of diagonal inter-chain couplings. The introduction of  
the diagonal hopping leads to the appearance of two separate transition points, giving rise to a mixed-phase zone, which is essential for the appearance 
of ME  in the energy spectrum~\cite{me1,me2}. In addition, energy-dependent MEs can be generated in 1D AAH models by incorporating 
long-range hopping~\cite{lr1,lr2}, short-range dimerized hopping~\cite{dim1, dim2}, or by engineering the structure of the 
quasiperiodic potentials\cite{qp1,qp2,qp3}.

Recently, the study of open quantum systems have gained significant attention, since realistic quantum systems are inherently 
coupled to their surrounding environments. Such systems play a pivotal role across diverse fields, including chemistry, atomic 
and molecular physics, quantum optics, condensed matter physics, quantum information, and quantum  computation
~\cite{os1,os2,os3,os4,os5}. Open quantum systems can be described using a non-Hermitian Hamiltonian, where physical gain 
and loss are captured by introducing imaginary terms into the Hamiltonian, and their properties are analyzed through the resulting 
complex eigenvalue spectrum~\cite{osnh1,osnh2,osnh3,osnh4}. The non-Hermitian systems can show exotic phenomena such 
as non-Hermitian skin effect~\cite{se1,se2,se3,se4}, parity-time ($\mathcal{PT}$) phase transition~\cite{pts1,pts2,pts3}, and 
exceptional points~\cite{ep,ep1,ep2}. Moreover, open systems have been realized experimentally in various systems, including optical, 
mechanical, and electrical  setups, which highlight the practical significance of non-Hermitian systems~\cite{exp1,exp2,exp3,exp4,exp5}. 
Anderson localization has been explored in disordered optical lattice systems, revealing that the presence of physical gain 
and loss can enhance the localization of light~\cite{and1,and2}. In addition, the role of $\mathcal{PT}$-symmetry in the non-Hermitian 
Aubry-André model has also received considerable attention~\cite{ptaah1,ptaah2}. Despite significant progress, non-Hermitian 
Hamiltonians only  describe the dynamics at short times, whereas the long-time, unconditional evolution is governed by the Liouvillian 
of the master equation~\cite{liouv1,liouv2,liouv3}. Although localization has been extensively studied in closed systems, and 
recent works have been addressed environmental effects through non-Hermitian Hamiltonians by introducing imaginary terms, a 
comprehensive understanding of localization in open quasiperiodic systems is still lacking. Exploring a Liouvillian-based framework 
is essential to reveal how both coherent dynamics and dissipation reshape localization transitions in realistic scenario.

To address this issue, we investigate the effect of environmental dissipation on localization properties of a quasiperiodic 
two stranded ladder. In our approach, the essential features of non-Hermitian quasicrystals are incorporated into the dissipative 
quantum jump processes described by a Lindblad master equation. Within this framework, the long-time dynamical transitions are 
directly linked to the phase transitions of the corresponding non-Hermitian effective Hamiltonian. We introduce single-site Lindblad 
operators in two different configurations, applied to every site in one case and to alternating sites in the other, in order to examine the 
impact of distinct dissipation patterns. Using third quantization and calculating the bath matrices corresponding to dissipation in the 
Majorana fermionic representation, we obtain the system’s damping matrix. From its eigenfunctions, we compute the inverse and 
normalized participation ratios to analyze the localization properties. Our results indicate that dissipation gives rise to a rich phase 
diagram with extended, mixed, and localized phases, where the mixed-phase window and critical point are tunable through the 
dissipation strength. Specifically, when dissipation acts on every sites under balanced gain-loss conditions, the localization 
transition occurs at a critical point same as closed AAH ladder and there is no mixed-phase zone. Introducing imbalance leads to the 
emergence of a mixed-phase zone, which broaden as imbalance increases. In contrast, when dissipation is applied to alternating sites, 
a mixed-phase zone arises even under balanced condition. In both the scenarios, the critical values of localization transition decreases with 
increasing imbalance between gain and loss. Interestingly, our results indicate that in presence of dissipation, a mixed-phase zone can 
emerge in a one-dimensional AAH ladder even in absence of diagonal hopping or without imposing any constraints on the quasiperiodic potential.

\begin{figure}[ht]
	{\centering\resizebox*{6cm}{3.5cm}{\includegraphics{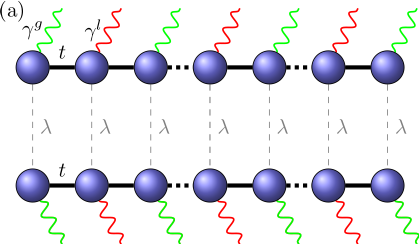}}\\
		\resizebox*{6cm}{3.5cm}{\includegraphics{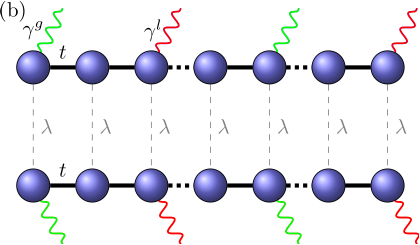}}}
	\caption{(Color online). Schematic representation of two-stranded Aubry-André-Harper ladders illustrating two dissipation configurations: 
		(a) alternating gain and loss at every site, and (b) alternating gain and loss applied at every other site. $\gamma^g$ and $\gamma^l$ denote 
		the strengths of gain and loss dissipation, respectively, while $t$ and $\lambda$ represent the intra-chain and inter-chain hopping amplitudes, 
		respectively.}
	\label{model}
\end{figure}

The rest of the paper is structured as follows: Sec. II introduces the model, Hamiltonian, and the theoretical framework based on third 
quatization formalism within Majorana fermionic representation. In Sec. III, we present and analyze the results in detail. Finally, Sec. IV 
provides the conclusions.

\section{Model and Methods} 
We consider a open quantum system of fermions in dissipative environment described by the Lindblad master equation
~\cite{LB1,LB2}
\begin{equation}
\frac{d\rho}{dt} = \hat{\mathcal{L}}\rho = -i[H, \rho] 
+ \sum_{m} \left( L_{m}\rho L_{m}^{\dagger} 
- \frac{1}{2}\{L_{m}^{\dagger}L_{m}, \rho\} \right),
\label{equ1}
\end{equation}
where $\rho$ is the density matrix, $H$ is the system's Hamiltonian. $\hat{\mathcal{L}}$ denotes the Liouvillian 
superoperator and $L_m$ are Lindblad operators represent the effect of different baths. From a quantum-trajectory 
viewpoint, in the absence of the quantum-jump contributions $L_{m}\rho L_{m}^{\dagger}$, the short-time evolution is 
governed by the effective non-Hermitian Hamiltonian $H_{\text{eff}} = H_{0} - i \sum_{m} L_{m}^{\dagger} L_{m}$
~\cite{LB3,LB4}. Conversely, the long-time evolution is determined by the Liouvillian superoperator $\mathcal{L}$, 
with the steady-state density matrix $\rho_s$ satisfying $\mathcal{L}\rho_s=0$. 

If all $L_m$ are linear in fermionic operator, the master equation generates quadratic Lindbladians. By applying third 
quantization approach~\cite{thirdQ1,thirdQ2}, these quadratic Lindbbladians can be diagonalized in the canonical basis 
of Majorana fermions. A single complex fermion can always be represented in terms of a pair of Majorana fermions, for 
instance
$$
w_{2j-1} = (c_j + c_j^{\dagger}), \quad 
w_{2j} = i(c_j - c_j^\dagger) \, ,
$$
satisfying the anti-commutation relation 
$$
\{ w_{j}, w_{k} \} = 2 \delta_{j,k}, 
\qquad j,k = 1,2,\ldots,2n.
$$
Now the system's Hamiltonian and the bath operators can be written in the Majorana fermion basis as

\begin{equation}
H = \sum_{i,j} H_{ij} w_{i} w_{j},
\label{equ2}
\end{equation}
\begin{equation}
L_{m} = \sum_{j} l_{m,j} w_{j}.
\label{equ3}
\end{equation}
The Hilbert space is now mapped onto the $2^{2n}$-dimensional Liouville space $\mathcal{K}$, which is spanned by the new 
set of Majorana operators $P_{\alpha} = w_{1}^{\alpha_{1}}w_{2}^{\alpha_{2}}\cdots w_{2n}^{\alpha_{2n}}, $ where 
$\alpha_{j} \in \{0,1\}.$ In this canonical basis the Liouvillian of Eq.~\eqref{equ1} contain both the even and odd parity 
subspaces and it can be written in the following bilinear form
\begin{equation}
\hat{\mathcal{L}} = \frac{1}{2} \sum_{i,j} 
\begin{pmatrix}
\hat{c}_i^{\dagger} & \hat{c}_i
\end{pmatrix}
A_{ij}
\begin{pmatrix}
\hat{c}_j \\
\hat{c}_j^{\dagger}
\end{pmatrix}
- A_0 \hat{\mathbf{I}} \, ,
\label{equ4}
\end{equation}
This equation needs to define in two parity subspaces. Since, a physical observable contain even number of Majorana fermions 
operator, we can consider this equation in even subspaces only, in which the structure matrix $\mathbf{A}$ of dimension 
$4N\times4N$ takes the form~\cite{thirdQ3}
\begin{equation}
	\mathbf{A} =
	\begin{pmatrix}
	- \mathbf{X}^\dagger & -i \mathbf{Y} \\
	0 & \mathbf{X}
	\end{pmatrix}.
	\label{equ5}
\end{equation}
with
\begin{eqnarray}
\mathbf{X} = -4i \mathbf{H} + 2 \left( \mathbf{M} + \mathbf{M}^T \right),\nonumber\\
\mathbf{Y} = -4i \left( \mathbf{M} - \mathbf{M}^T \right),\nonumber\\
\mathbf{A_0} = \tfrac{1}{2} \mathrm{Tr}\,\mathbf{X}.
\label{equ6}
\end{eqnarray}
M is a complex Hermitian matrix which provides the parametrization for the Lindblad operators
\begin{equation}
	M_{ij} = \sum_{m} l_{m,i} l^{*}_{m,j}.
\label{equ7}
\end{equation}
The Hamiltonian matrix $\mathbf{H}$ and the bath matrix $\mathbf{M}$, both have the dimensions $4N\times4N$. Since, 
$\mathbf{A}$ has a block-triangular form, the spectrum of $\hat{\mathcal{L}}$ is fully determined by the eigenvalues 
$\beta_i$ of matrix $\mathbf{X}$, called the rapidities. By calculating eigenvectors of $\mathbf{A}$, we can construct 
the normal master modes $b_j$, $b_j^\prime$ and the Liouvillian takes the diagonal form 
$\hat{\mathcal{L}}=\sum_{j}\beta_j b_j^\prime b_j$. The eigenspectrum of $X$, contain all the information about liouvillian 
spectrum. It can be shown that, the eigenspectrum of the damping matrix and that of the non-Hermitian effective Hamiltonian 
contains the same information~\cite{damping1}.

As shown in Fig.~\ref{model}, our system consist of a quasiperiodic two-stranded ladder with local baths having $N$ lattice 
sites in each strand. We consider two different dissipation configurations: in the first, local gain and loss are applied 
alternately on every site, while in the second, they are applied alternatively on every other site. The tight-binding Hamiltonian of the 
ladder is given by

\begin{eqnarray}
H & = & \sum_{p=\text {I,II}}\Big[\sum_{j=1}^{N} \epsilon_{p,j} c_{p,j}^{\dagger} c_{p,j} \nonumber \\
& + & \sum_{j=1}^{N-1} t_p\Big(c_{p,j}^{\dagger} c_{p,j+1} + c_{p,j+1}^{\dagger}c_{p,j}\Big)\Big] \nonumber \\
& + & \sum_{j=1}^{N}\lambda \Big(c_{\tiny\mbox{I},j}^{\dagger} c_{\tiny\mbox{II},j} + c_{\tiny\mbox{II},j}^{\dagger} c_{\tiny\mbox{I},j}\Big)
\label{equ8}.
\end{eqnarray}
Here, $p$ ($=\small\mbox{I},\small\mbox{II}$) labels the strand index, while $j$ denotes the lattice sites along each strand. 
The operators $c_{p,j}^\dagger$ and $c_{p,j}$ are the standard fermionic creation and annihilation operators, respectively. 
The parameter $t_p$ represents the intra-strand hopping amplitude for strand $p$, and $\lambda$ denotes the inter-strand coupling strength.

The parameter $\epsilon_{p,j}$ represents the on-site energy of an electron at site $j$ of the strand $p$. Under the influence of AAH 
modulation, the on-site energies along both strands can be written as
\begin{equation}
\epsilon_{\tiny\mbox{I},j}=\epsilon_{\tiny\mbox{II},j}= V \cos \left(2 \pi b j \right)
\label{equ9}
\end{equation}
where $b=(\sqrt{5}-1)/2$.

For the bath, we choose single-site lindblad operators for gain and loss as:
\begin{eqnarray}
L_{p,j}^g=\sqrt{\gamma^g} c_{p,j}^\dagger; \nonumber \\ 
L_{p,j}^l=\sqrt{\gamma^l} c_{p,j}; 
\label{equ10}
\end{eqnarray}
where $\gamma^g$ and $\gamma^l$ are the gain and loss strength respectively.

Now in Majorana fermion basis $w_{p,2j-1} = (c_{p,j} + c_{p,j}^{\dagger});\quad w_{p,2j} = i(c_{p,j} - c_{p,j}^\dagger)$ 
the Hamiltonian of Eq.~\eqref{equ8} and the Lindblad operators of Eq.~\eqref{equ10}, can be written as

\begin{eqnarray}
H & = & \sum_{p=\text {I,II}}\Big[\frac{1}{2}\sum_{j=1}^{N} \epsilon_{p,j} (1+ i w_{p,2j} w_{p,2j-1}) \nonumber \\
& + & \frac{i t_p}{2} \sum_{j=1}^{N-1} \Big(w_{p,2j} w_{p,2j+1} - w_{p,2j-1}w_{p,2j+2}\Big)\Big] \nonumber \\
& + & \frac{i \lambda}{2} \sum_{j=1}^{N} \Big(w_{\tiny\mbox{I},2j} w_{\tiny\mbox{II},2j-1} - w_{\tiny\mbox{I},2j-1} w_{\tiny\mbox{II},2j}\Big)
\label{equ11}
\end{eqnarray}
\begin{eqnarray}
L_{p,j}^g=\frac{\sqrt{\gamma^g}}{2} \big(w_{p,2j-1}+ i w_{p,2j}\big) \nonumber\\ 
L_{p,j}^l=\frac{\sqrt{\gamma^l}}{2} \big(w_{p,2j-1}- i w_{p,2j}\big).
\label{equ12}
\end{eqnarray}

We construct the bath matrix $M$ using Eq.~\eqref{equ7} and Eq.~\eqref{equ12} in Majorana fermion basis. 
For dissipation at every site, the bath matrix takes the form
\[
M=\frac{1}{4}
\begin{pmatrix}
\gamma^g & -i\gamma^g & 0  &  0    &  0    &  0 & 0 & 0 \\
i\gamma^g & \gamma^g &  0 &   0   &   0  &    0 & 0 & 0 \\
0 & 0  &\gamma^l  &  i\gamma^l   & 0 &  0 & 0 & 0    \\
0 & 0 & -i\gamma^l& \gamma^l & 0 & 0 & 0 & 0       \\
&      &      & \ddots &      &      \\
0&   0   &  0    &  0    & \gamma^g & -i\gamma^g & 0 & 0 \\
0&   0   &  0    &  0   & i\gamma_g & \gamma_g & 0 & 0\\
0&   0   &  0    &  0   & 0 & 0 & \gamma^l & \gamma^l \\
0&   0   &  0    &  0   & 0 & 0 & -i\gamma^l & \gamma^l \\
\end{pmatrix}.
\]
For dissipation applied at every other site, the bath matrix is given by
\[
M=\frac{1}{4}
\begin{pmatrix}
\gamma^g & -i\gamma^g & 0  &  0    &  0    &  0 & 0 & 0 \\
i\gamma^g & \gamma^g &  0 &   0   &   0  &    0 & 0 & 0 \\
0 & 0  & 0  &  0   & 0 &  0 & 0 & 0    \\
0 & 0 & 0 & 0 & 0 & 0 & 0 & 0       \\
&      &      & \ddots &      &      \\
0&   0   &  0    &  0    & \gamma^l & i\gamma^l & 0 & 0 \\
0&   0   &  0    &  0   & -i\gamma^l & \gamma^l & 0 & 0\\
0&   0   &  0    &  0   & 0 & 0 & 0 & 0 \\
0&   0   &  0    &  0   & 0 & 0 & 0 & 0\\
\end{pmatrix}
\]
From Eq.~\eqref{equ6}, the corresponding damping matrix 
$X$ can then be obtained by using the above bath matrices.
\begin{figure*}[ht]
	{\centering\resizebox*{17cm}{4cm}{\includegraphics{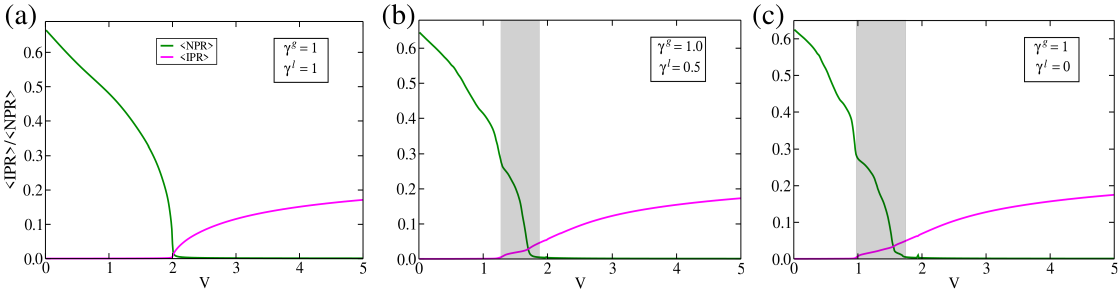}}}
	\caption{(Color online).  $\langle\mathrm{IPR}\rangle$ and $\langle\mathrm{NPR}\rangle$ as functions of the Aubry-André-Harper 
		disorder strength, with dissipation applied at every site. Panels (a)–(c) correspond to: (a) $\gamma^g=1,\ \gamma^l=1$; 
		(b) $\gamma^g=1,\ \gamma^l=0.5$; and (c) $\gamma^g=1,\ \gamma^l=0$. Here the system size is $2N=3000$.}
	\label{iprnpr}
\end{figure*}
\begin{figure*}[ht]
	{\centering\resizebox*{15cm}{5cm}{\includegraphics{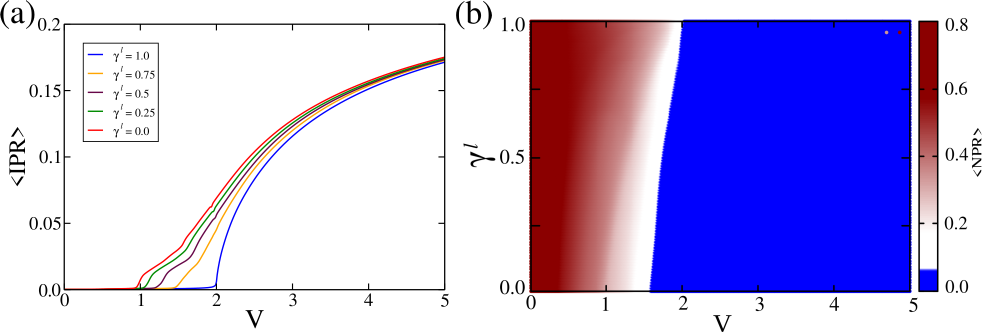}}}
	\caption{(Color online).  With dissipation applied at every site:
		(a) $\langle$IPR$\rangle$ as a function of the Aubry-André-Harper disorder strength $V$ for $\gamma^l = 1, 0.75, 0.5, 0.25,$ and $0$, 
		with fixed $\gamma^g = 1$ and system size $2N = 3000$.
		(b) Phase diagram in the parameter space spanned by $V$ and $\gamma^l$, where the color bar represents the $\langle$NPR$\rangle$. 
		Here $\gamma^g = 1$ and the system size is $2N = 500$.}
	\label{iprnprglw}
\end{figure*}
To see the impact of dissipation on localization transitions, we analyze the properties of the eigenstates of the 
damping matrix. If $|\psi_n\rangle$ is the normalized eigenstate of $n$th state and $4N$ is the dimension of the 
damping matrix $X$, then the degree of localization can be quantified using inverse participation ratio (IPR) and 
normalized participation ratio (NPR) defined as 
\begin{eqnarray}
\text{IPR}_n = \frac{\sum_{j=1}^{4N} \left|\psi_{n,j}\right|^{4}}
{\left(\langle \psi_n | \psi_n \rangle \right)^{2}}
\label{equ13}
\end{eqnarray}
and
\begin{eqnarray}
\text{NPR}_n=(4N \times \text{IPR}_n)^{-1}
\label{equ14}
\end{eqnarray}
In thermodynamic limit, for a delocalized eigenstate IPR$=0$ and NPR$\ne 0$, whereas for a localized eigenstate, 
IPR$\ne 0$ and NPR$=0$. To gain a comprehensive understanding of the spectrum, we evaluate the average IPR and NPR 
by summing over all eigenstates, defined as follows.
\begin{equation}
\langle\text{IPR}\rangle = \frac{1}{4N} \sum_{n=1}^{4N} \text{IPR}_n
\label{equ15}
\end{equation}
and
\begin{equation}
\langle\text{NPR}\rangle = \frac{1}{4N} \sum_{n=1}^{4N} \text{NPR}_n
\label{equ16}
\end{equation}

\section{Numerical results and discussion}
\begin{figure*}[ht]
	{\centering\resizebox*{17cm}{4cm}{\includegraphics{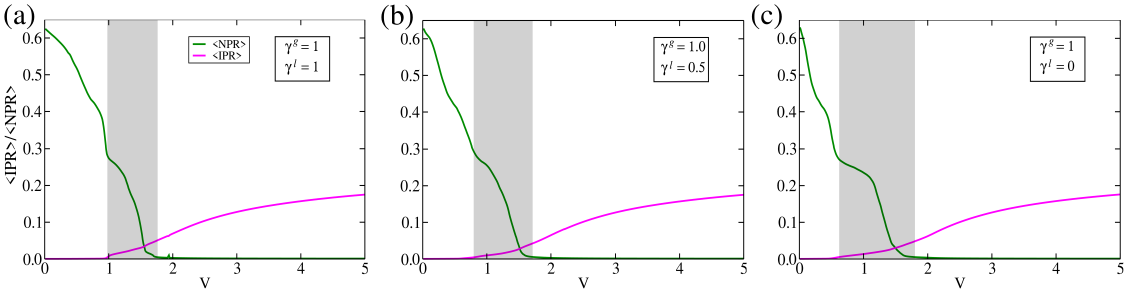}}}
	\caption{(Color online). $\langle\mathrm{IPR}\rangle$ and $\langle\mathrm{NPR}\rangle$ as functions of the Aubry-André-Harper 
		disorder strength, with dissipation applied at alternate site. Panels (a)–(c) correspond to: (a) $\gamma^g=1,\ \gamma^l=1$; 
		(b) $\gamma^g=1,\ \gamma^l=0.5$; and (c) $\gamma^g=1,\ \gamma^l=0$. Here the system size is $2N=3000$. }
	\label{aiprnpr}
\end{figure*}
\begin{figure*}[ht]
	{\centering\resizebox*{15cm}{5cm}{\includegraphics{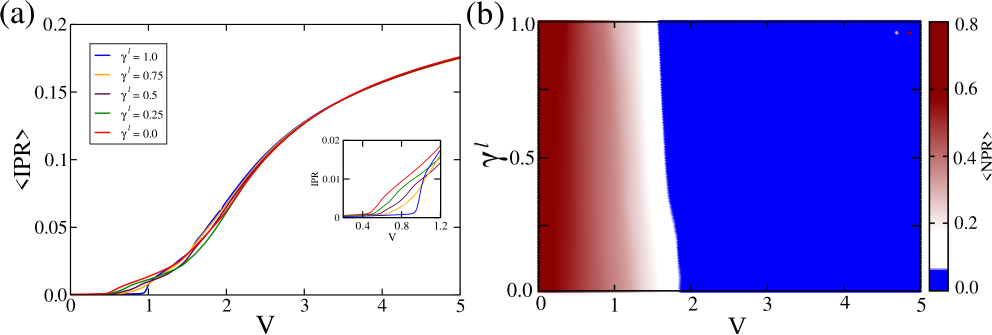}}}
	\caption{(Color online). With dissipation applied at alternate site:
		(a) $\langle$IPR$\rangle$ as a function of the Aubry-André-Harper disorder strength $V$ for $\gamma^l = 1, 0.75, 0.5, 0.25,$ and $0$, 
		with fixed $\gamma^g = 1$ and system size $2N = 3000$.
		(b) Phase diagram in the parameter space spanned by $V$ and $\gamma^l$, where the color bar represents the $\langle$NPR$\rangle$. 
		Here $\gamma^g = 1$ and the system size is $2N = 500$.}
	\label{aiprnprglw}
\end{figure*}

Based on the above theoretical prescription now we present our essential results. The main focus is to discuss
the intricate interplay between dissipation and quasiperiodicity on localization transition of AAH ladder. We will 
consider two configurations: (i) dissipation at every site, and (ii) dissipation at alternate sites. The balanced and imbalanced 
condition can be tuned by varying the relative strengths of loss and gain. For both these cases we critically 
analyze the effect of dissipation strength on critical point and mixed-phase zone. The values of parameters which are 
kept constant throughout the calculations are: intra-chain hopping $t_{\tiny\mbox{I}}=t_{\tiny\mbox{II}}=1$, inter-strand
hopping $\lambda=1$, and unless stated otherwise each strand consists of a total of $N=1500$ lattice sites.

\subsection{Dissipation at every site}
In this case, each site is subject to dissipation, with the Lindblad operators acting alternately as gain and loss. 
Specifically, the gain and loss operators are acting as:
\begin{align}
L_{p,j}^g=\sqrt{\gamma^g} c_{p,j}^\dagger; \qquad \text{for odd j}\nonumber \\ 
L_{p,j}^l=\sqrt{\gamma^l} c_{p,j}; \qquad \text{for even j}\nonumber
\end{align}
When $\gamma^g = \gamma^l$, the system is in balanced gain-loss condition, while $\gamma^g \neq \gamma^l$ corresponds to 
an imbalanced regime.
In Fig.~\ref{iprnpr}, we plot average inverse participation ratio ($\langle \text{IPR} \rangle$) and normalized participation ratio 
($\langle\text{NPR}\rangle$) as a function of disorder strength, $V$. Figure~\ref{iprnpr}(a) represents the balanced gain-loss condition
 where $\gamma^g=\gamma^l=1$. Here we can see that, a transition from a delocalized phase to a localized phase occurs at critical 
 point $V=2$. No mixed-phase region is observed, since there is no parameter range where both the $\langle$IPR$\rangle$ and $\langle$NPR$\rangle$ are simultaneously 
 nonzero. This result is similar to a isolated AAH ladder where the transition occurs at $V/t=2$~\cite{me2}. Now we introduce an imbalance in the 
 system by taking the gain and loss dissipation strength as $\gamma^g=1$ and $\gamma^l=0.5$ which is shown in Fig.~\ref{iprnpr}(b).
 This result indicate a delocalization to localization transition through a mixed-phase between $V=1.28$ and $V=1.90$ where
 both $\langle\text{IPR}\rangle$ and $\langle\text{NPR}\rangle$ are finite. In Fig.~\ref{iprnpr}(c), we show the result for maximum 
 imbalance condition where $\gamma^g=1$ and $\gamma^l=0$, i.e., only gain dissipation at alternate sites. Here mixed phase zone 
 appears in between $V=0.98$ and $V=1.87$ which indicates that mixed-phase window increases with increasing the imbalanced dissipation 
 in the system.
 
To obtain a complete picture of the delocalization to localization transition, In Fig.~\ref{iprnprglw} we plot $\langle\text{IPR}\rangle$ 
and $\langle$NPR$\rangle$ by changing $\gamma^l$ and keeping $\gamma^g=1$. Figure~\ref{iprnprglw}(a) shows the variation
of $\langle$IPR$\rangle$ with $V$ for $\gamma^l=1, 0.75, 0.5, 0.25, 0$ (blue, orrange, maroon, green, red). Here we can see that by decreasing 
$\gamma^l$ and hence increasing imbalance between loss and gain, the critical value for phase transition from  delocalized to mixed-phase 
zone gradually decreases, making the system easier to be localized. In Fig.~\ref{iprnprglw}(b), we display the $\langle$NPR$\rangle$ 
in the two-dimensional parameter space V versus $\gamma^l$, in which the color bar indicates the $\langle$NPR$\rangle$ values. 
Here we fix the system size $2N=500$ and the red, white, and blue color zones represent delocalized, mixed and localized regime, respectively. When gradually increasing 
$\gamma^l$, mixed-phase zone decreases and the transition point from the mixed phase to the localized phase moves toward larger disorder strengths. At $\gamma^l=1$, the transition occurs at $V/t=2$.

\subsection{Dissipation at alternate site}

In this configuration, the Lindblad operators act only on every other site, arranged 
alternately as gain and loss described by
\begin{align}
L_{p,j}^g=\sqrt{\gamma^g} c_{p,j}^\dagger; \qquad \text{for j}=1,5,9,\ldots\nonumber \\ 
L_{p,j}^l=\sqrt{\gamma^l} c_{p,j}; \qquad \text{for j}=3,7,11,\ldots\nonumber
\end{align}
In Fig.~\ref{aiprnpr}, we plot $\langle$IPR$\rangle$ and $\langle$NPR$\rangle$ as a function of $V$ for three 
different values of $\gamma^l$ keeping $\gamma^g=1$. The balanced condition ($\gamma^g=\gamma^l$=1)
is shown in Fig.~\ref{aiprnpr}(a). Unlike previous case, this result shows that mixed-phase zone appear in between 
$V=0.98$ and $V=1.78$ even in balanced condition. With increasing the imbalance between gain and loss by decreasing 
$\gamma^l$ to $0.5$ which is shown in Fig.~\ref{aiprnpr}(b), the mixed phase zone increases and lying between
$V=0.75$ and $V=1.8$. Figure~\ref{aiprnpr}(c) shows the result for maximum imbalance, with $\gamma^l=0$. In this 
case the transition from delocalization to localization phase occurs through mixed-phase window between 
$V=0.56$ and $V=1.86$. Here the width of the mixed-phase zone become maximum. 

Now, like the previous case, to have a complete understanding of the role of dissipation, here we also plot $\langle$IPR$\rangle$ 
for five different values of $\gamma^l$  and $\langle$NPR$\rangle$ for all possible values of $\gamma^l$ as a function of $V$
in Fig.~\ref{aiprnprglw}(a) and Fig.~\ref{aiprnprglw}(b) respectively.  From the inset of Fig.~\ref{aiprnprglw}(a) it is clear that,
with decreasing $\gamma^l$ i.e, increasing the imbalance the critical point for transition from delocalized  to mixed-phase zone
shifts gradually towards the lower disorder value. Figure~\ref{aiprnprglw}(b) shows the dependence of $\langle$NPR$\rangle$ on
$V$ and $\gamma^l$, where the $\langle$NPR$\rangle$ values are indicated by the color bar. Here we keep system size $2N=500$. Red, white, and blue regions represent
the delocalized, mixed, and localized regime respectively. This result indicates that the localization transition has a strong 
dependence on dissipation strength $\gamma^l$. With increasing $\gamma^l$ the width of the mixed-phase zone decreases and the transition 
from the mixed phase to the localized phase occurs towards the lower disorder value. At balanced condition,
i.e, for $\gamma^l=1$ the critical point of transition occur at $V/t=1.57$.


\section{closing remarks}
In this work, we have explored localization transition in a quasiperiodic ladder 
in presence of environmental dissipation. The quasiperiodicity is described by AAH model. 
Two different kinds of dissipation configurations have been taken into account. For one case,
alternate gain and loss are applied to each site and for other case it has applied to every 
other site. The central focus of this work is to explore the interplay between dissipation 
and disorder on localization transition. A tight-binding model is employed to construct 
the Hamiltonian of the ladder, while the effect of dissipation is included using the Lindblad master 
equation formalism. The third quantization method within the Majorana fermionic
representation has been used to calculate the damping matrix. The inverse and normalized
participation ratios has been obtained from the damping matrix to analyze the localization 
phenomena. The key results of our work are as follows.\\
(i) Dissipation induces a complex phase diagram comprising extended, mixed, and localized regimes
without imposing any constraints on the hopping parameter or quasiperiodic potential. The critical
points and mixed-phase zone can be tuned by varying the dissipation strength.\\
(ii) For dissipation acting alternatively on all sites under balanced gain-loss conditions, 
the localization transition occurs at the same critical point as in the closed AAH ladder, 
and no mixed-phase region is observed. However, introducing gain-loss imbalance results in 
the appearance of a mixed-phase region, which progressively broadens with increasing imbalance.\\
(iii) When alternating gain-loss dissipation is applied to every other sites, a mixed-phase region 
emerges even under perfectly balanced gain-loss conditions and the width of this mixed-phase zone increases
with increasing imbalance.\\
(iv) In both dissipation schemes, the critical disorder strength for the transition from the delocalized 
to the mixed phase shifts toward lower disorder values as the imbalance between gain and loss increases. 
In contrast, for the transition from the mixed to the localized phase, the critical disorder strength 
decreases with increasing imbalance in the case of dissipation at each site, whereas it increases for 
dissipation applied at every other site.

Our results reveal that dissipation can fundamentally reshape the localization properties of quasiperiodic AAH ladders, 
giving rise to tunable transitions between delocalized, mixed, and localized phases. Remarkably, such a mixed-phase regime 
can occur even without diagonal hopping or additional constraints on the quasiperiodic potential, highlighting dissipation 
as an effective control parameter for engineering phase transitions in non-Hermitian quasiperiodic systems.

\section*{ACKNOWLEDGMENTS}

SS is thankful to ANRF, India (File number: PDF/2023/000319) for providing her research fellowship. SS acknowledges JNCASR, India for funding.
SKP acknowledges the JC Bose fellowship and ANRF (File No. ANRF/JCB/SKP/4719), Govt. of India for the financial assistance. The authors
thank Dr. Subhasis Sinha for useful discussions.

\end{document}